\documentclass[structabstract]{aa}  
\usepackage{graphicx}
\usepackage{amssymb}
\usepackage{epstopdf}
\usepackage{booktabs} % for much better looking tables
\usepackage{float}
\usepackage{array} % for better arrays (eg matrices) in maths
\usepackage{paralist} % very flexible & customisable lists (eg. enumerate/itemize, etc.)
\usepackage{verbatim} 
\usepackage{amsmath}
\usepackage{graphicx}
\usepackage{graphicx}
\usepackage{booktabs} % for much better looking tables
\usepackage{array} % for better arrays (eg matrices) in maths
\usepackage{paralist} % very flexible & customisable lists (eg. enumerate/itemize, etc.)
\usepackage{verbatim} % adds environment for commenting out blocks of text & for better verbatim
\usepackage{subfig} % make it possible to include more than one captioned figure/table in a single float
\usepackage{graphicx}
\usepackage{amssymb}
\usepackage{epstopdf}
\usepackage[english]{babel}

\def\be{\begin{equation}}
\def\ee{\end{equation}}
\def\bea{\begin{eqnarray}}
\def\eea{\end{eqnarray}}

\usepackage{txfonts}
\usepackage{longtable,lscape}
\usepackage{longtable}

\begin{document}
%
  % \title{Controlling systematic effects with a new calibration method in bolometric interferometry: the self-calibration}
   \title{Self-calibration: an efficient method to control systematic effects in bolometric interferometry}

   %\subtitle{I. Overviewing the $\kappa$-mechanism}

%   \author{G. Wuchterl
%          \inst{1}
 %         \and
%%          C. Ptolemy\inst{2}\fnmsep\thanks{Just to show the usage
 %         of the elements in the author field}
 %         }

 \author{M.-A. Bigot-Sazy \inst{1}  \and R. Charlassier \inst{2} \and J.-Ch. Hamilton \inst{1} \and J. Kaplan \inst{1} \and G. Zahariade \inst{1}}
 
 \institute{APC, Astroparticule et Cosmologie, Universit\'e Paris Diderot, B\^atiment Condorcet, 10 rue Alice Domon et L\'eonie Duquet, F-75205, Paris Cedex 13, France  \and Turing-Solutions, 25 rue Dauphine, 75006 Paris, France}
 
%\email{mabigot@apc.univ-paris7.fr}  
  %\author{V.?Arsenijevic\inst{\ref{inst1}}\and S.?Fabbro\inst{\ref{inst2}}\and A.?M.?Mour\?ao\inst{\ref{inst3}}\and A.?J.?Rica da Silva\inst{\ref{inst1}}}
  
  %            T\"urkenschanzstrasse 17, A-1180 Vienna\\
    %          \email{wuchterl@amok.ast.univie.ac.at}
    %     \and
    %         University of Alexandria, Department of Geography, ...\\
    %         \email{c.ptolemy@hipparch.uheaven.space}
      %       \thanks{The university of heaven temporarily does not
      %               accept e-mails}
     %        }

 \date{Received September 21, 2012}
%  \date{Received September 15, 1996; accepted March 16, 1997}
\offprints{M.-A. Bigot-Sazy, \email{mabigot@apc.univ-paris7.fr}}
% \abstract{}{}{}{}{} 
% 5 {} token are mandatory
 
  \abstract
  % context heading (optional)
  % {} leave it empty if necessary  
   {The QUBIC collaboration is building a bolometric interferometer dedicated to the detection of B-mode
polarization fluctuations in the Cosmic Microwave Background.}
  % aims heading (mandatory)
   {We introduce a self-calibration procedure Ð related to those used in radio-interferometry to control a large range of
instrumental systematic errors in polarization-sensitive instruments.}
  % methods heading (mandatory)
   {This procedure takes advantage of the fact that in the absence of systematic effects, measurements on redundant baselines should exactly match each other. For a given systematic error model, measuring each baseline independently therefore allows to write a system of nonlinear equations whose unknowns are the systematic error model parameters (gains and couplings of Jones matrices for instance).}
  % results heading (mandatory)
   {We give the mathematical basis of the self-calibration. We implement this method numerically in the context of bolometric interferometry. We show that, for large enough arrays of horns, the nonlinear system can be
solved numerically using a standard nonlinear least-squares fitting and that the accuracy achievable on systematic effects is only limited by the time spent on the calibration mode for each baseline apart from the validity of the systematic error model.}
  % conclusions heading (optional), leave it empty if necessary 
   {}


   \maketitle
%
%________________________________________________________________

\section{Introduction}
The quest for the B-mode of the polarization of the cosmic background is one of the scientific priorities of observational cosmology today. Observing this mode appears as the most powerful way to constrain inflation models. However, detecting such a weak signal is a real experimental challenge. In addition to a high statistical sensitivity (huge number of horns and bolometers required), future experiments will need excellent quality of foreground removal and unprecedented control of instrumental effects. 

Currently, most projects are based on the experimental concept of an imager. A promising alternative technology is bolometric interferometry. This is the project of the QUBIC instrument (the QUBIC collaboration 2010). A first module is planned to be installed at the Franco-Italian Concordia Station in Dome C, Antarctica in 2014. The aim is to combine the advantages of an imager in terms of sensitivity with those of an interferometer in terms of control of systematic effects. The statistical sensitivity of the QUBIC instrument is comparable to that of an imager with the same number of horns covering the same sky fraction. The full QUBIC instrument (six modules) will comprise three frequencies (97, 150 and 220 GHz) and aims to constrain, at the 90\% confidence level, a tensor-to-scalar ratio of 0.01 with one year of data.

The aim of this article is to introduce a new method, specific to bolometric interferometry called self-calibration, and to give an example of application with the QUBIC instrument. This method allows to control a wide range of instrumental systematic effects. 

This self-calibration technique  is based on the redundancy of the receiver array (Wieringa 1991). It uses the fact that in the absence of systematic effects, redundant baselines of the interferometer should measure exactly the same quantity. For a real instrument, these measurements will be different because of systematics. These small differences can be used to calibrate parameters that characterize completely the instrument for each channel and estimate the instrumental errors.

In the case of a bolometric interferometer, the square horn array will provide a large number of redundant baselines. In this way, a bolometric interferometer can be self-calibrated thanks to a calibration mode during which it will measure separately the $n_h(n_h-1)/2$ baselines or a fraction of the $n_h(n_h-1)/2$ baselines with $n_h$ the number of horns observing an external polarized source.

This method is inspired by traditional interferometry (Pearson \& Readhead 1984) where signal phases are often lost due to atmospheric turbulence. Standard calibration procedures, similar to those used for imaging techniques and based on observations of an unresolved source whose flux is assumed to be known, exist in radio-interferometry. We emphasize that the word self-calibration refers, by opposition, to a procedure in which no knowledge of the observed source is required (most of the time, the object which is scientifically studied is itself used as the calibration source). Most of these self-calibration techniques are based on the evaluation of so-called closure quantities Ð phases or amplitudes. A set of unknown phases can for instance be iteratively reconstructed by forming quantities where they are nulled (the product of the three visibilities that can be formed with three antennas). The use of redundant baselines for calibration is by contrast rather uncommon in radio-interferometry $-$ see (Wieringa 1991; Noordam \& de Bruyn 1982). This comes from the fact that most radio-interferometers have very few redundant baselines; in order to achieve very high angular resolution, it is indeed better to arrange a given number of antennas in order to optimize the uv-plane sampling, rather than to maximize redundancy.

A new kind of all-digital radio-interferometer the ``omniscope'', dedicated to 21 cm observations, has recently been proposed by (Tegmark \& Zaldarriaga 2009a,b); its concept can be summarized by the following five steps: 

1) signals collected by antennas are digitized right after amplification, 

2) a temporal Fast Fourier Transform (FFT) is performed in order to split them into frequency subbands, 

3) a spatial FFT is performed on each subband set, 

4) the square modulus of the FFT result is computed, 

5) an inverse spatial FFT is performed to recover the visibilities. 

There are conceptual similarities with the bolometric interferometer concepts, since in this latter instrument, steps 2 and 3 are performed in an analogical way respectively by the beam combiner and the bolometers. Because antennas have to be located on a grid in order for the FFTs to be performed, an omniscope will possess many redundant baselines; (Liu et al. 2010) have shown how this allows to self-calibrate the antenna complex gains. In both cases (standard radio-interferometer or omniscope), the aim is to calibrate the complex antenna gains; there is actually a mathematical trick to get a linear system of equations from which these gains can be obtained. 

This paper is organized as follows. Section 2 introduces the self-calibration method with the Jones matrix formalism and the Mueller matrix formalism in the case of radio-interferometry and omniscope. We show in section 3 how the procedure can be applied to the QUBIC bolometric interferometer and finally, we describe the possible self-calibration algorithm and its results.
%__________________________________________________________________
\section{General principle}
%__________________________________________________________________
\subsection{Instrumental systematics modelisation with Jones matrices}
In this section, we use the notation proposed by (O'Dea et al. 2007). With the electric field of an incident radiation at a frequency $\phi$ defined as $\mathbb{R}[{\overrightarrow{E}e^{-i\phi t}}]$, and choosing two basis vectors $\overrightarrow{e_x}$ and $\overrightarrow{e_y}$ orthogonal to the direction of propagation $\overrightarrow{k}$, all the statistical information is encoded in the coherence matrix $\mathbf{C}$

\bea
%\begin{equation}
\hspace{2cm} \mathbf{C}=
\begin{pmatrix}
C_{xx} & C_{xy}  \\ 
C_{xy}^* & C_{yy}
\end{pmatrix}
=
\begin{pmatrix}
\langle\mid E_x \mid^2\rangle  & \langle E_xE_y^{*} \rangle \\ 
\langle E_x^{*}E_y \rangle  & \langle\mid E_y \mid^2\rangle
\end{pmatrix}  \nonumber \\
=\frac{1}{2}
\begin{pmatrix}
I+Q & U-iV \\ 
U+iV & I-Q
\end{pmatrix},
%\end{equation}
\label{eq:eq1}
\eea

where $E_x$, $E_y$ are complex amplitudes of the transverse electric field $\overrightarrow{E}$  and I, Q, U and V are the Stokes parameters.
\hspace{1cm}

The propagation of a incident radiation $\overrightarrow{E}$ through a receiver can be described by a Jones matrix $\mathbf{J}$ such that the electric field after passing through the receiver $\overrightarrow{E_r}$ is 
\begin{equation}
\overrightarrow{E_r}=\mathbf{J}\overrightarrow{E}.
\label{eq:eq2}
\end{equation}
where the Jones matrix $\mathbf{J}$ is a $2\times2$ complex matrix. It describes how the instrument linearly transforms the two dimensional vector representing the incoming radiation field $\overrightarrow{E}$ into the two dimensional vector of the outgoing field $\overrightarrow{E_r}$.

For an instrument with several components, the Jones matrix is the product of the Jones matrices for each component. 
For example, the ideal Jones matrix for an instrument in which the incident radiation passing through a rotating half-wave plate before propagating through the horns is

\bea
\hspace{2cm}  \mathbf{J}_{hhwp} & = & \mathbf{J}_{rot}^T\mathbf{J}_{hwp}\mathbf{J}_{rot}\mathbf{J}_{horn} \nonumber \\
& & =\begin{pmatrix}
cos(\omega t) & -sin(\omega t)  \\ 
sin(\omega t) & cos(\omega t)
\end{pmatrix} 
\begin{pmatrix}
1 & 0  \\ 
0 & -1
\end{pmatrix} \nonumber \\
&  & \times  \begin{pmatrix}
cos(\omega t) & sin(\omega t)  \\ 
-sin(\omega t) & cos(\omega t)
\end{pmatrix}
\begin{pmatrix}
1 & 0  \\ 
0 & 1
\end{pmatrix} \nonumber \\
&  & = \begin{pmatrix}
cos(2\omega t) & sin(2\omega t)  \\ 
sin(2\omega t) & -cos(2\omega t)
\end{pmatrix} ,
\label{eq:eq3}
\eea

where $\omega$ is the angular velocity of the half-wave plate,  $\mathbf{J}_{rot}$ is the rotation matrix,  $\mathbf{J}_{hwp}$ is the ideal Jones matrix of the half-wave plate and $\mathbf{J}_{horn}$ is the ideal Jones matrix of one horn.
 
 After passing through the receiver, ideal orthogonal linear detectors measure the power in two components
$$S_1 = \frac{1}{2}(I+Qcos(4\omega t)+Usin(4\omega t)) $$
$$S_2 = \frac{1}{2}(I-Qcos(4\omega t)-Usin(4\omega t)).$$

To model systematic errors within a polarization-sensitive interferometer, the Jones matrix can be described by introducing diagonal terms: the complex gain parameters $g_{x}$ and $g_{y}$ and non-diagonal terms: the complex coupling parameters $e_{x}$ and $e_{y}$ associated to the orthogonal polarizations
\begin{equation}
\mathbf{J}=
\begin{pmatrix}
1-g_{x} & e_{x}  \\
e_{y} & 1-g_{y} 
\end{pmatrix}.
\label{eq:eq4}
\end{equation}

Systematic errors arising from the half-wave plate and from the square horn array can be modeled by :
\begin{enumerate}
  \item 
a Jones matrix for the half-wave plate
\begin{equation}
\mathbf{J}_{hwp}=
\begin{pmatrix}
1-h_{x} & \xi_{x}  \\
\xi_{y} & -(1+h_{y}) 
\end{pmatrix} 
\label{eq:eq5},
\end{equation}
\item 
a Jones matrix for the horn $i$ with $1 \leqslant i\leqslant n_h$
\begin{equation}
\mathbf{J}_{horn,i}=
\begin{pmatrix}
1-g_{x,i} & e_{x,i}  \\
e_{y,i} & 1-g_{y,i} 
\end{pmatrix}.
\label{eq:eq6}
\end{equation}

\end{enumerate}

The electric field $\overrightarrow{E}$ propagated through  the half-wave plate and the horn $i$  becomes
\begin{equation}
\overrightarrow{E_i} =\mathbf{J}_{rot}^T\mathbf{J}_{hwp}\mathbf{J}_{rot}\mathbf{J}_{horn,i}\overrightarrow{E} .
\label{eq:eq7}
\end{equation}

\subsection{Instrumental systematics modelisation with Mueller matrix}

The Jones matrix expresses the transformation of the electric field in the x and y-directions and the Mueller matrix describes how the different polarization states transform. The Jones matrix is a $2 \times 2$ matrix whereas the Mueller matrix is a $4 \times 4$ matrix. So, the $4 \times 4$ Mueller matrix can be written as the direct product of the $2 \times 2$ Jones matrices. 

Let us calculate the tensor product of the outgoing field given by Eq.(\ref{eq:eq2})
\begin{equation}
\overrightarrow{E}_r ^*\otimes \overrightarrow{E}_r=\mathbf{J}\overrightarrow{E}  \otimes  \mathbf{J}^*\overrightarrow{E}^*.
\label{eq:eq8}
 \end{equation}
 
In general, the direct product $\langle \overrightarrow{E}  \otimes \overrightarrow{E}^* \rangle$ gives the vector $\overrightarrow{C}$ defined as

\begin{equation}
\overrightarrow{C}=\begin{bmatrix}
C_{xx}\\
C_{xy} \\
C_{yx}\\
C_{yy}\\
\end{bmatrix}
= \langle \overrightarrow{E}  \otimes \overrightarrow{E}^* \rangle=\langle \begin{bmatrix}E_x \\ E_y \end{bmatrix}
\otimes
\begin{bmatrix}E_x^* \\ E_y^* \end{bmatrix} \rangle=\begin{bmatrix} \langle E_xE_x^* \rangle \\ \langle E_xE_y^* \rangle   \\ \langle E_yE_x^* \rangle \\ \langle E_yE_y^* \rangle \end{bmatrix}.
\label{eq:eq9}
 \end{equation}
 Using Eq.(\ref{eq:eq8}), one can write the transmission of the electric field through an instrument described by its Jones matrix using the vector $\overrightarrow{C}$ of this electric field 
\begin{equation}
\overrightarrow{C}_{r}=\mathbf{J}\overrightarrow{E}  \otimes  \mathbf{J}^*\overrightarrow{E}^*=\mathbf{J} \otimes  \mathbf{J}^*\overrightarrow{C}.
\label{eq:eq10}
\end{equation}

Accordingly Eq.(\ref{eq:eq1}), the polarization state of this electric field can be described by the Stokes vector $\overrightarrow{S}$ defined by
\begin{equation}
\overrightarrow{S}=\begin{bmatrix}
I \\
Q \\
U \\
V \\
\end{bmatrix}
=
\begin{bmatrix}
C_{xx}+C_{yy} \\
C_{xx}-C_{yy} \\
C_{xy}+C_{yx} \\
i(C_{xy}-C_{yx}) \\
\end{bmatrix}.
\label{eq:eq11}
\end{equation}

where I, Q, U and V are the Stokes parameters.

One can obtain the expression of the Stokes vector from the vector $\overrightarrow{C}$ defined in Eq.(9)
\begin{equation}
\overrightarrow{S}=\mathbf{A}\overrightarrow{C}
\label{eq:eq12},
\end{equation}
where \hspace{2.5cm}  $\mathbf{A}=
\begin{bmatrix}
1 & 0 & 0 & 1\\
1 & 0 & 0 & -1 \\
0 & 1 & 1 & 0\\
0 & i & -i & 0\\
\end{bmatrix}.
$
\vspace{0.5cm}

Substituting Eq.(\ref{eq:eq12}) into Eq.(\ref{eq:eq10}), it follows that the outgoing Stokes vector $\overrightarrow{S}_r$ can be written as 
\begin{equation}
\overrightarrow{S}_r=\mathbf{A}(\mathbf{J} \otimes  \mathbf{J}^*)\mathbf{A}^{-1}\overrightarrow{S}=\mathbf{M}\overrightarrow{S}
\label{eq:eq13}
\end{equation}

where $\mathbf{M}=\mathbf{A}(\mathbf{J} \otimes  \mathbf{J}^*)\mathbf{A}^{-1}$ is the Mueller matrix which describes how the Stokes parameters transform.

\subsection{Polarized measurement equation with Mueller formalism}

A polarization-sensitive interferometer measures the complex Stokes visibilities from all baselines defined by the horns i and j in an array of receivers
\begin{equation}
V_{ij}=\begin{pmatrix}
V_{ij}^I \\
V_{ij}^Q \\
V_{ij}^U \\
V_{ij}^V \\
\end{pmatrix}.
\label{eq:eq14}
\end{equation}

These vectors could reduce to a scalar or a vector with 2, 3 or 4 elements depending on the Stokes parameters the instrument is sensitive to. One can define $a= {1,2,3,4}$ the number of Stokes parameters the instrument allows to measure.

The $n_h (n_h-1)/2$ baselines of the interferometer can be classified into $n_\neq$ sets $s_\beta$ of redundant baselines (same length, same direction) indexed by $\beta$.
In the absence of systematic errors, the redundant visibilities should have exactly the same values
\begin{equation}
\forall\begin{Bmatrix}i,j\end{Bmatrix}\epsilon s_\beta ,   V_{ij}=V_\beta.
\label{eq:eq15}
\end{equation}

For a real instrument however, redundant visibilites $\bar{V}_{ij}$ will not have exactly the same values because of systematic errors and statistical (photon) noise, and one can write the following system of $a\times n_h (n_h-1)/2$ complex equations
\begin{equation}
\bar{V}_{ij}=\mathbf{M}_{ij}.V_\beta+n_{ij}
\label{eq:eq16}
\end{equation}

where $n_{ij}$ are statistical noise terms and where $\mathbf{M}_{ij}$ are a kind of complex Mueller matrices which reduce to the identity matrix for a perfect instrument.
\begin{equation}
\mathbf{M}_{ij}=\begin{pmatrix}
M_{ij}^{II}&M_{ij}^{IQ}&M_{ij}^{IU}&M_{ij}^{IV}\\
M_{ij}^{QI}&M_{ij}^{QQ}&M_{ij}^{QU}&M_{ij}^{QV}\\
M_{ij}^{UI}&M_{ij}^{UQ}&M_{ij}^{UU}&M_{ij}^{UV}\\
M_{ij}^{VI}&M_{ij}^{VQ}&M_{ij}^{VU}&M_{ij}^{VV}\\
\end{pmatrix}.
\label{eq:eq17}
\end{equation}
The elements of these matrices are not independent and can be expressed in terms of the diagonal and the non-diagonal terms of the Jones matrix. 

The first order Mueller matrix for a polarization sensitive experiment is
\begin{equation}
M_{ij}=\frac{1}{2}\begin{pmatrix}A&B \\ C&D\end{pmatrix}
\end{equation}
with 
\begin{equation}
A=\begin{pmatrix}2+g^*_{x,j}+g^*_{y,j}+g_{x,i}+g_{y,i} & g^*_{x,j}-g^*_{y,j}+g_{x,i}-g_{y,i}\\ g^*_{x,j}-g^*_{y,j}+g_{x,i}-g_{y,i} & 2+g^*_{x,j}+g^*_{y,j}+g_{x,i}+g_{y,i} \end{pmatrix}
\end{equation}
\begin{equation}
B=\begin{pmatrix}e^*_{x,j}+e^*_{y,j}+e_{x,i}+e_{y,i} & -i(e^*_{x,j}-e^*_{y,j}-e_{x,i}+e_{y,i})\\ e^*_{x,j}-e^*_{y,j}+e_{x,i}-e_{y,i} & -i(e^*_{x,j}+e^*_{y,j}-e_{x,i}-e_{y,i}) \end{pmatrix}
\end{equation}
\begin{equation}
C=\begin{pmatrix}e^*_{x,j}+e^*_{y,j}+e_{x,i}+e_{y,i} & -e^*_{x,j}+e^*_{y,j}-e_{x,i}+e_{y,i}\\ -i(e^*_{x,j}-e^*_{y,j}-e_{x,i}+e_{y,i}) & i(e^*_{x,j}+e^*_{y,j}-e_{x,i}-e_{y,i}) \end{pmatrix}
\end{equation}
\begin{equation}
D=\begin{pmatrix}2+g^*_{x,j}+g^*_{y,j}+g_{x,i}+g_{y,i} & i(g^*_{x,j}-g^*_{y,j}-g_{x,i}+g_{y,i})\\ -i(g^*_{x,j}-g^*_{y,j}+g_{x,i}+g_{y,i}) & 2+g^*_{x,j}+g^*_{y,j}+g_{x,i}+g_{y,i} \end{pmatrix}.
\end{equation}

\subsection{Standard radio-interferometry and omniscope}
In the case of instruments that are not designed to measure polarized radiation fields (like most radio-interferometers or omniscopes), the Mueller matrix reduces to its first element $(a = 1)$,
\begin{equation}
\mathbf{M}_{ij} \rightarrow M_{ij}^{II}=(1+g_i)(1+g_j^*)
\label{eq:eq23}
\end{equation}
where $g_i$, $g_j$ are the complex gains of antennas $i$ and $j$. A linear system can be obtained by taking the complex logarithm of Eq.(\ref{eq:eq16})
\begin{equation}
v_{i,j} =v_\beta+G_i+G^*_j
\label{eq:eq24}
\end{equation}
where $v_{i,j}=log\begin{vmatrix}v_{i,j}^I\end{vmatrix}$, $v_{\beta}=log\begin{vmatrix}v_{\beta}^I\end{vmatrix}$, $G_i=log\begin{vmatrix}g_i\end{vmatrix}$ and $G^*_j=log\begin{vmatrix}g_j^*\end{vmatrix}$.
In order to get an invertible system, one must add an overall gain normalization constraint
\begin{equation}
\sum_{i=0}^{n_h-1}G_i=0
\label{eq:eq25}.
\end{equation}
For a square grid array of $n_h\gtrsim 8$ antennas, the problem becomes overdetermined since the number of equations is $2 \times n_h(n_h-1)/2\sim n_h^2$   with $n_h(n_h-1)/2$ the number of the measured complex visiblilties. The number of unknowns is $2n_\neq+2n_h\sim6n_h$ with $n_\neq$ the number of complex visibilities of the unique baselines and $n_h$ the number of complex gain factors.
Things are however made subtler by the use of the complex logarithm $-$ see (Liu et al. 2010) for details.

 \subsection{Polarization-sensitive interferometer}
 In the polarization-sensitive case, on the contrary, Eq.(\ref{eq:eq16}) cannot be easily transformed into a linear system. The number of unknowns is $a\times2n_\neq+ 8n_h$, while the number of equations is still $\propto$ $n_h^2$. In order to get an invertible system, one must add the gain normalization constraint given by Eq.(\ref{eq:eq25}) and another overall normalization constraint for the non-diagonal terms of the Jones matrix as the leakages are small and randomly distributed
\begin{equation}
\sum_{i=0}^{n_h-1}e_i=0.
\label{eq:eq26}
\end{equation}
The system (which becomes overdetermined for $n_h \gtrsim 20$ if $a =3$) can then be solved using a standard nonlinear least-squares method.

\section{Application to the QUBIC bolometric interferometer}

 \subsection{Observables in bolometric interferometry}
 In this section, we derive the expression for the power received in the focal plane in the case of bolometric interferometry.
 
The bolometric interferometer proposed with the QUBIC instrument (the QUBIC collaboration 2010) is the millimetric equivalent of the first interferometer dedicated to astronomy: the Fizeau interferometer.

The receptors are two arrays of $n_h$ horns: the primary and secondary horns back-to-back on a square grid behind the optical window of a cryostat.
Filters and switches are placed in front and between the horn array. The switches will be used only during the calibration phase. The polarization of the incoming field is modulated using a half-wave plate located before the primary horns. This location of the half-wave plate avoids a leakage from the Stokes parameter I to the Stokes parameters Q an U if the half-wave plate has no inhomogeneities.

Signals are correlated together using an optical combiner.
The interference fringe patterns arising from all pairs of horns, with a given angle, are focused to a single point on the focal plane. Finally, a polarizing grid splits the signal into x and y-polarizations, each being focused on a focal plane equipped with bolometers. These bolometers measure a linear combination of the Stokes parameters modulated by the rotating half-wave plate.

With an interferometer, the correlation between two receivers allows for direct access to the Fourier modes (visibilities) of the Stokes parameters I, Q and U. In the case of a bolometric interferometer, the observable is the superposition of the fringes formed by the sky electric field passing through a large number of back-to-back horns and then focused on the detector plane array. The image on the focal plane of the optical combiner is the synthesized image as only specific Fourier modes are selected by the receiving horns array. A bolometric interferometer is therefore a synthesized imager whose beam is the synthesized beam formed by the array of receiving horns.

The electric field $\bar{E}^\eta_{iq}(\hat{n}_p)$ collected by the horn $i$ located at $ \overrightarrow{x_i}$ for the polarization $\eta \in \left\{x,y\right\}$ and measured by a bolometer $q$ when all the primary horns are looking at the same radiation field $E ^{\eta}(\hat{n}_p)$ as a function of the observed direction $\hat{n}_p$ is
\begin{equation}
\begin{pmatrix}
\bar{E}^x_{iq}(\hat{n}_p) \\
\bar{E}^y_{iq}(\hat{n}_p)
\end{pmatrix}
=\mathbf{\alpha}_{iq}\mathbf{\beta}_{q}(\hat{n}_p)
\begin{pmatrix}
E^x(\hat{n}_p) \\
E^y(\hat{n}_p)
\end{pmatrix}.
\label{eq:eq27}
\end{equation}

where we have introduced the matrices $\mathbf{\alpha}_{iq}$ and $\mathbf{\beta}_{i}(\hat{n}_p)$ which characterize completely the instrument.

A matrix $\mathbf{\alpha}_{iq}$ is defined for each channel of horns (i) and bolometers (q) and includes the geometrical phases induced by the beam combiner,  the beams of the secondary horns $B_{sec}(\hat{d_q})$ and the gain $g_q$  of the bolometer

\begin{equation}
\mathbf{\alpha}_{iq}^{ideal}=
\begin{pmatrix}
\alpha_{iq}^{x} & 0 \\ 
0 & \alpha_{iq}^{y}
\end{pmatrix}
\label{eq:eq28}
\end{equation}
with $$\alpha_{iq}^{x}=g_q \int B^x_{sec}(\hat{d}_q)exp[i2\pi\frac{\overrightarrow{x}_i}{\lambda}\hat{d}_q]J(\nu)\Theta(\hat{d}-\hat{d}_q)d\nu d\hat{d}_q$$
and, $$\alpha_{iq}^{y}=g_q \int B^y_{sec}(\hat{d}_q)exp[i2\pi\frac{\overrightarrow{x}_i}{\lambda}\hat{d}_q]J(\nu)\Theta(\hat{d}-\hat{d}_q)d\nu d\hat{d}_q .$$

The detector location is given by the unit vector $\hat{d}_q$ and $\lambda$ is the wavelength of the instrument.  The integrations are over the surface of each individual bolometer modeled with the top-hat like function $\Theta(\hat{d}-\hat{d}_q)$ and on bandwidth of the instrument $J(\nu)$ with $\nu$ the frequency.

A matrix $\mathbf{\beta}_i(\hat{n}_p)$ is defined for each channel of pointings (p) and horns (i) and includes the primary beam $B_{prim,i}^\eta(\hat{n}_p)$, the horn position $\overrightarrow{x_i}$ and the direction of pointing $\hat{n}_p$ 
\begin{equation}
\mathbf{\beta}_{i}^{ideal}(\hat{n}_p)=
\begin{pmatrix}
\beta_{i}^{x}(\hat{n}_p) & 0 \\ 
0 & \beta_{i}^{y}(\hat{n}_p)
\end{pmatrix}
\label{eq:eq29}
\end{equation}
with $$ \beta_{i}^{x}(\hat{n}_p)=B_{prim,i}^x(\hat{n}_p)exp(i2\pi\overrightarrow{x_i}\hat{n}_p) $$
and,
$$ \beta_{i}^{y}(\hat{n}_p)=B_{prim,i}^y(\hat{n}_p)exp(i2\pi\overrightarrow{x_i}\hat{n}_p) .$$

The half-wave plate rotates at angular speed $\omega$ and therefore modulates the two orthogonal polarizations such that
\begin{equation}
\begin{pmatrix}
E'^x \\ 
E'^y
\end{pmatrix}
=
\begin{pmatrix}
cos(2\omega t) & sin(2\omega t)\\ 
sin(2\omega t) & -cos(2\omega t)
\end{pmatrix}.\begin{pmatrix}
E^x \\ 
E^y
\end{pmatrix}.
\label{eq:eq30}
\end{equation}

The electric field collected by one horn $i$ located at $\overrightarrow{x_i}$ is
\begin{equation}
s_i^{\eta}(\hat{n}_p)=E'^\eta(\hat{n}_p)\mathbf{\beta}_{i}(\hat{n}_p).
\label{eq:eq31}
\end{equation}
The electric field collected by one horn $i$ located at $\overrightarrow{x_i}$ and reaching the bolometer $q$ located at $\hat{d}_q$ is
\begin{equation}
s_{iq}^{\eta}(\hat{n}_p)=E'^\eta(\hat{n}_p)\mathbf{\beta}_{i}(\hat{n}_p)\mathbf{\alpha}_{iq}.
\label{eq:eq32}
\end{equation}

The power measured by one polarized bolometer located at $\hat{d}_q$ in the focal plane of the beam combiner is then

\begin{equation}
S_{qp}^{\eta} = \int \left| E'^\eta(\hat{n}_p) \right| ^2 \left| B_{q,s}(\hat{n}_p)\right|^2 d\hat{n}_p
\label{eq:eq33}
\end{equation}
where the synthesized beam $B_{q,s}(\hat{n}_p)$ for the detector q is formed by the arrangement of the primary horn array as follows
\begin{equation}
B_{q,s}(\hat{n}_p)=\sum_{i\leqslant j}\mathbf{\alpha}_{iq}\mathbf{\alpha}_{jq}^*\mathbf{\beta}_{i}(\hat{n}_p)\mathbf{\beta}_{j}^*(\hat{n}_p).
\label{eq:eq34}
\end{equation}
The synthesized beam depends on the sky direction $\hat{n}_p$.  So, the synthesized image is a convolution of the sky and of the electric field through the synthesized beam.

One can rewrite Eq.(\ref{eq:eq33}) to exhibit the modulation of the polarization induced by the half-wave plate:
\begin{equation}
  S^{\eta}_{qp}(t)=S_{qp}^I +\epsilon^{\eta}cos(4 \pi \omega t)S_{qp}^Q +\epsilon^{\eta}sin(4 \pi \omega t)S _{qp}^U
  \label{eq:eq35}
\end{equation}
where $\epsilon^x=1$ for the polarization $x$,  $\epsilon^y=-1$ for the polarization $y$ and where $S_{qp}^X$ are the synthesized images on the focal plane for each Stokes parameter $X=\left\{I, Q, U\right\}$. The cosine and sine coefficients come from the modulation induced by the rotating half-wave plate.

Systematic effects arising at any level of the detection can be modeled by associating a Jones matrix to each horn i $\mathbf{J}_{horn,i}$ and a Jones matrix of the half-wave plate $\mathbf{J}_{hwp}$ .

They can be introduced as defined in Eq.(\ref{eq:eq7}).

 \subsection{Self-calibration procedure}
During the self-calibration mode, distinct from the ordinary data taking mode, the instrument scans a polarized source and measures one by one the $n_h (n_h-1)/2$ synthesized images from all baselines or only a fraction of them. In the QUBIC design, this can be achieved using switches located between the back-to-back horns. The switches are used as shutters that are operated independently for all channels and are only required during the calibration phase. One can modulate on/off a single pair of horns while leaving all the others open in order to access the synthesized images measured by each pair of horns alone.

This procedure requires the knowledge of the individual primary beams of each horn. The maps of the primary beams can be obtained independently through scanning an external unpolarized source.
 
By repeating this with all baselines, all bolometers and all directions of pointing, one can construct a system of equations whose unknowns are
 \begin{enumerate}
\item 
the complex coefficients  $\alpha_{iq}^\eta$, defined for each horn $i$, each bolometer $q$ and for each polarization $\eta$ which correspond to $4 n_h n_q$ parameters,
\item 
the horn location $\overrightarrow{x_i}$ ($2 n_h$ parameters),
\item 
 the direction of pointing $\hat{n}_p$ ($2 n_p$ parameters),
\item 
the complex horns systematic effects $g_{x,i}$, $g_{y,i}$, $e_ {x,i}$, $e_{y,i}$ defined for each horn $i$ and for each polarization $\eta$ ($8 n_h$ parameters),
\item 
the complex half-wave plate systematic effects $h_x$, $h_y,$ $\xi_x,$ $\xi_y$ and for each polarization $\eta$ ($4 n_h$ parameters).

 \end{enumerate}
 
For an instrument with $n_h$ horns, $n_q$ bolometers and for a scan of $n_p$ pointings, the number of unknows is
\begin{equation}
n_u = 4  \times n_h \times n_q + 2\times n_h + 2 \times n_p + 8\times n_h+8.
  \label{eq:eq36}
\end{equation}
 
During the self-calibration procedure, the number of constraints is given by the measurements, i.e. the synthesized images. One has the $n_h (n_h-1)/2 $ measured synthesized images for each bolometer $n_q$, each pointing $n_p$, each Stokes parameter and each focal plane. With Eq.(\ref{eq:eq25}) and Eq.(\ref{eq:eq26}), one can add four complex equations : the sum of the diagonal and the non-diagonal terms of the Jones matrices. The number of constraints is given by
\begin{equation}
n_c = 6  \times n_h \times (n_h-1)/2 \times n_q  \times n_p + 8 .
  \label{eq:eq37}
\end{equation}

The problem becomes easily overdetermined: for an instrument with $n_h=9$, $n_q=4$ and $n_p=10$, the number of constraints is $8648$ and the number of unknows is $262$. 
It can be solved with a least squares algorithm.

The first module of the QUBIC instrument will consist in 400 primary horns and two 1024 element bolometer arrays. The number of constraints could be reduced if the self-calibration is not performed on the $n_h (n_h-1)/2$ baselines but on a fraction of baselines. This will be shown in the following.

However, closing all the switches except two would actually change dramatically the thermal load on the cryostat, which could affect the bolometric measurements. Fortunately, there is a trick explained in Appendix A that allows to indirectly measure $S_{ijpq}^\eta(\hat{d}_q)$ while minimaly changing the thermal load. One can show that
 \begin{equation}
S_{ijpq}^\eta(\hat{d}_q)= \bar{C}_{ipq}^\eta(\hat{d}_q)+\bar{C}_{jpq}^\eta(\hat{d}_q)-2\bar{S}_{ijpq}^\eta(\hat{d}_q)
  \label{eq:eq38}
 \end{equation}
where $\bar{S}_{ijpq}^\eta(\hat{d}_q)$ is the quantity measured by a bolometer $q$ when all the switches are open except the $i$ and $j$, and $\bar{C}_{ipq}^\eta(\hat{d}_q)$, $\bar{C}_{jpq}^\eta(\hat{d}_q)$ are the powers measured when all the switches are open except respectively i or j. Measuring these three terms therefore allows to measure $S_{ijpq}^\eta(\hat{d}_q)$ while keeping the thermal load almost constant. 

However, this also increases the noise. The noise on each term is therefore
\begin{multline*}
\Delta S^2_{ijpq} = \Delta \bar{C}^2_{ipq} + \Delta \bar{C}^2_{jpq}+4\Delta\bar{S}^2_{ijpq} \\
 %=2\times(n_h-1)\times\frac{NET^2}{T ^2} +4\times(n_h-2)\times \frac{NET^2}{T ^2} \nonumber \\
 =2 (n_h-1)NET^2 +4(n_h-2) NET^2 \\
    = (6n_h-10)NET^2
\end{multline*}
where $NET$ is the noise equivalent temperature of the bolometers and $T$ the temperature of the 100\% polarized source.

 \subsection{Numerical simulation} 
We have numerically implemented the method to check if the nonlinear system could be solved 

We generate the instrument with a set of ideal parameters (horn locations, directions of pointing, primary and secondary beams, detector locations...) and a set of parameters randomly corrupted by systematic errors (horn location errors, pointing errors, assymetries of beams, bolometer location errors, diagonal and non-diagonal terms of the Jones matrices ...). The widths of the random deviation of all corrupted parameters around their ideal value are given in Table 1.

In order to get a solvable system, one must add some normalization constraints for the coefficients $\alpha_{iq}^{\eta}$ and $\beta_i^{\eta}(\hat{n}_p)$ which do not change the modelling of systematic errors. They mean that the self-calibration only allows for relative calibration of these parameters. One can add:
 \begin{enumerate}
\item
an absolute calibration of the global gain of the instrument,  $\alpha_{00}^{\eta}=1$.
\item
a convention on the phase of the $\alpha_{iq}^{\eta}$  coefficients, $\forall q, arg(\alpha_{0q}^{\eta})=0$.

A rotation of global phase $\phi_q$ applied to the coefficients $\alpha_{iq}^{\eta}$ for one bolometer $q$ does not modify Eq.(\ref{eq:eq27}) and Eq.(\ref{eq:eq34}) and therefore the observations.
\item
a convention on the primary beams, $\forall i, \beta_{i}^{\eta}(\hat{0}))=1$

Multiplying the $\alpha_{iq}^{\eta}$  coefficients by a term defined as $c_ie^{\phi_i}$ and dividing the $\beta_i^{\eta}(\hat{n}_p)$  coefficients by the same term does not modify Eq.(\ref{eq:eq27}) and Eq.(\ref{eq:eq34}) and therefore the observations.

\item
a convention on the phase of the $\beta_i^{\eta}(\hat{n}_p)$  coefficients, $\forall p, arg(\beta_{0}^{\eta}(\hat{n}_p))=0$.

A rotation of global phase $\phi(\hat{n}_p)$ applied to the $\beta_i^{\eta}(\hat{n}_p)$ coefficients  for one pointing $p$ does not modify Eq.(\ref{eq:eq27}) and Eq.(\ref{eq:eq34}) and the observations.
 \end{enumerate}

 \vspace{0.5cm}  

We compute the corrupted synthesized images and add gaussian statistic noise given by
\begin{equation}
n_{noise}=\frac{NET_P}{\sqrt{t_b} \sqrt{2} } \sqrt{6n_h-10} 
  \label{eq:eq39}
\end{equation}

where the noise equivalent temperature of the bolometers NET is taken to be $300\mu K.s^{\frac{1}{2}}$, the temperature of the 100\% polarized source is T=100K and the time spent on each baseline on the calibration mode is $t_b= 1$s. The usual convention is to give the NET for unpolarized detectors but it is convenient in our case to use quantities with polarization. In this case, the NET is given by $NET=\frac{NET_P}{\sqrt{2}}$.

 \vspace{0.5cm}  

We solve the non-linear system with a standard non-linear least-squares method based on a Levenberg-Marquardt algorithm. The ideal coefficients (without systematic errors) are used as starting guess for the different parameters.

\begin{table}[h] 
\centering
\begin{tabular}{c r} \hline\hline Error source&\multicolumn{1}{c}{Gaussian level} \\ [0.5ex] 
\hline	% inserts single-line 
Pointing Uncertainty $\hat{n}_p$ & 1 {[deg]} \\	% Entering row contents 
Horn location error $\overrightarrow{x_i}$& 100 {[$\mu$m]} \\ 
Detector location error $\hat{d}_q$& 1 {[$\mu$m]} \\
Bolometer gain $g_q$ & 0.01\\
Primary beam error $B^{\eta}_{prim,i}(\hat{n}_p)$ & 0.01\\
Secondary beam error $B^{\eta}_{sec}(\hat{d}_q)$ & 0.01 \\ 
Horn systematics $g_{{\eta},i}$ & $0.0001$ \\
Horn systematics $e_{{\eta},i}$ & $0.0001$ \\
Half-wave plate systematics $h_{\eta}$ & $0.01$ \\
Half-wave plate systematics $\xi_{\eta}$ & $0.01$ \\[1ex]
% centering table % creating eight columns %inserting double-line
\hline \end{tabular} 
\caption{ \textit{Range for systematic errors for each parameter.}}  %title of the table
\label{tab:hresult} 
\end{table}

\begin{figure*}
\centering\includegraphics[width=19.2cm]{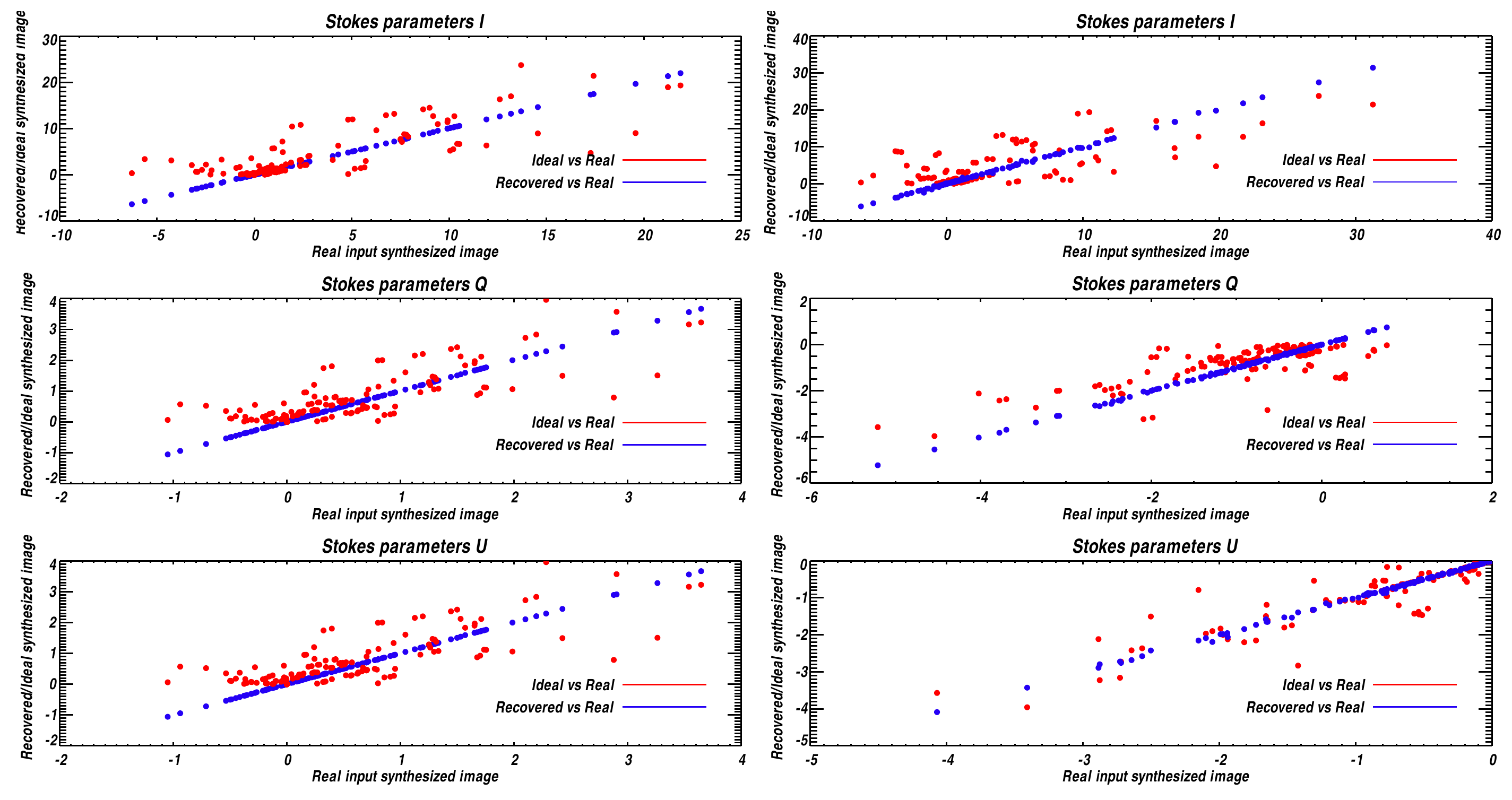}
\caption{\textit{Results of the self-calibration simulation for the synthesized images $S_{qp}^{I}$ $S_{qp}^{Q}$ and $S_{qp}^{U}$ for the X focal plane at right and Y focal plane at left for an instrument with 9 primary horns, 9 bolometers and 10 pointings for a time spent on calibration mode for each baseline of $t_b=1s$ and 100 realisations. These plots represent scatter plots of ideal vs. real synthesized image in red and of recovered vs. real in blue. The six plots show the advantage of the self-calibration method. }
\label{fig:radish}}
\end{figure*}

\subsection{Results}

In order to be able to have a large number of realizations, we run the simulation for an array of 9 primary horns, 9 bolometers and 10 pointings for 100 realisations.
Fig.1 shows the result of the self-calibration simulation. The six plots are scatter plots of ideal vs. real synthesized images in red and of recovered vs. real synthesized images in blue for the X and Y focal planes and each Stokes parameter I, Q and U. The synthesized images are computed with Eq.(33), the ideal synthesized images are the synthesized images without systematic effects, the real synthesized images are the simulated measurements and the recovered synthesized images are computed with the output parameters of the self-calibration simulation. 
The red plots show the corruption after adding the systematic effects defined in Table 1. The blue plots show that the corruption is solved after applying the self-calibration method. 

This method allows access to the systematic effects of the horns and of the half-wave plate.
It also allows to calibrate the parameters $\alpha_{iq}^\eta$ and $\beta_{i}(\hat{n}_p)$ that characterize completely the instrument for each channel of pointings, horns, bolometers.

%\begin{figure*}[h]
%\centering\includegraphics[width=19.2cm]{/Users/marie-annebigot-sazy/Desktop/result_selfcal_polarization/papier2/reconstruction.pdf}
%\caption{\textit{Results of the self-calibration simulation for the synthesized images $S_{qp}^{I}$ $S_{qp}^{Q}$ and $S_{qp}^{U}$ for the X focal plane at right and Y focal plane at left for an instrument with 9 primary horns, 9 %bolometers and 10 pointings for a time spent on calibration mode for each baseline of $t_b=1s$ and 100 realisations. These plots represent scatter plots of ideal vs. real synthesized image in red and of recovered vs. real in %blue. The six plots shows the advantage of the self-calibration method. }
%\label{fig:radish}}
%\end{figure*}

%\begin{figure}
%\centering\includegraphics[width=9cm]{/Users/marie-annebigot-sazy/Desktop/result_selfcal_polarization/papier2/reconstrucyty.pdf}
%\title{\textit{Y Focal plane}}
%\caption{\textit{Results of the self-calibration simulation for the synthesized images $S_{qp}^{I}$ $S_{qp}^{Q}$ and $S_{qp}^{U}$ for the Y focal plane for an instrument with 9 primary horns, 9 bolometers and 10 pointings for a time spent on calibration mode for each baseline of $t_b=1s$ and 100 realisations. These plots represents a scatter plot of ideal vs. real synthesized image in red and of recovered vs. real in blue. The three plots shows the advantage of the self-calibration method.The three plots shows the same result as Fig.1.}
%\label{fig:radish}}
%\end{figure}

In running the simulation, one can find that the residual error on each output parameter will depend on the number of horns, bolometers, pointings, baselines per pointings and on the time spent measuring each baseline.
Adding more horns, pointings, bolometers and baselines increases the mathematical constraints on a given measurement, it allows to form new baselines and adds redundancy on the horn array.
Fig.2 shows that the residual diagonal term error of the Jones matrix of the half-wave plate improves as the number of baselines per pointing is higher. This result was obtained for a simulation with 9 primary horns, 9 bolometers and 10 pointings for 40 realisations and for a time spent on each baseline $t_b=1s$. The baselines measured for each pointing are chosen randomly. Similar plots are obtained with the other parameters defined in Table 2.

%\begin{figure}[H]
%\centering\includegraphics[width=8cm]{/Users/marie-annebigot-sazy/Desktop/result_selfcal_polarization/polarisation_script_hwp_calshist/polarisation/nbbsperpointing_gain3.pdf}
%\caption{\textit{Results of the self-calibration simulation for the diagonal terms of the Jones matrix of the half-wave plate and a time spent per baseline $t_b=1s$. This plot represents the residual error on these parameters as a function of the percent of baselines per pointing. The red line represents a power law of the shape $\sim \frac{c}{n_{bs}^\Phi}$ with c a constant, $n_{bs}$  the number of baselines per pointing and $\Phi$ the index of the power law given Table 2. This law gives the limit of the relative accuracy that can be achieved on the systematic parameters. The green line represents the input error on the systematic effects of the half-wave plate in the simulation given Table 1.}
%\label{fig:radish}}
%\end{figure}

One can put together these variables and define a power law which allows to calculate the residual error on each parameter defined in Table 1:
\begin{equation}
\sigma_{c} = c \times \frac{1}{t_b^k} \times \frac{1}{n_h^\alpha} \times \frac{1}{n_b^\beta} \times \frac{1}{n_p^\gamma} \times \frac{1}{n_{bs}^\Phi}
  \label{eq:eq40}
\end{equation}
with $c$ a constant, k the  exponent of the measuring time per baseline $t_b$ and $\alpha$, $\beta$, $\gamma$, $\Phi$ the exponent of the number of horns $n_h$, bolometers $n_q$, pointings $n_p$ and the percent of baselines per pointing $n_{bs}$. 

The values for each index are summarized in Table 2 for two different measuring times per baseline $t_b=1$  and $t_b=100s$.

\begin{table*}[ht] 
\centering
\begin{tabular}{| c | c | c | c | c | c | c | c | c | c | c | }
\hline
$$& \multicolumn{4}{c|}{$t_b=1s$} & \multicolumn{4}{c|}{$t_b=100s$}  \\
\hline
 parameters& $\alpha$ & $\beta$ & $\gamma$ & $\Phi$ & $\alpha$ & $\beta$ & $\gamma$ & $\Phi$\\
\hline
 $\alpha_{iq}^\eta$  & 0.74& 0.99& 0.75 & 0.84& 0.97& 1.55 & 1.22&1.06\\
 $\hat{n}_p$  & 0.46 &0.37& 0.24 & 0.73 & 0.75& 0.70& 1.02 &1.09\\
 $\overrightarrow{x_i}$  & 0.58& 0.66& 0.97 & 0.82& 0.76 & 1.04& 1.28 &1.18\\
 $g_\eta(\overrightarrow{x_i})$& 0.91& 1.06& 0.83 & 0.48& 1.04& 1.52 & 0.45 &0.75\\
 $e_\eta(\overrightarrow{x_i})$ & 0.77& 1.18& 0.36 & 0.65& 1.01& 1.16& 0.63 &0.99\\
$h_\eta $ & 0.55& 0.67& 0.12 & 0.59& 0.78& 0.85  & 0.25 &0.78\\
 $\xi_\eta$  & 0.64& 0.58& 0.11 &0.38& 1.11&0.84&0.46&0.80\\ 
\hline
\end{tabular}
\caption{\textit{Results of the self-calibration simulation for each recovered parameter given in the first column for an instrument with 9 horns, 9 bolometers and 10 pointings. Following the power law given by Eq.(\ref{eq:eq40}), one can calculate the exponent for each variable: the exponent $\alpha$ for the number of horns, the exponent $\beta$ for the number of bolometers, the exponent $\gamma$ for the number of pointings and $\Phi$ the number of baselines per pointing. This work has been done for two different measuring times per baseline $t_b=1s$ and $t_b=100s$ and with 40 realisations of the simulation.}}
\end{table*}

One can observe in Table 2 that the error on the different reconstructed parameters is better when the time spent on each baseline is longer.  Fig.3 represents the residual half-wave plate gain error as a function of the time spent on each baseline during the calibration mode.
It shows that the limitation of the accuracy achieved on the systematic parameters is given by the time spent on calibration mode for each baseline $t_b$.

\begin{figure}[!h]
\centering\includegraphics[width=8cm]{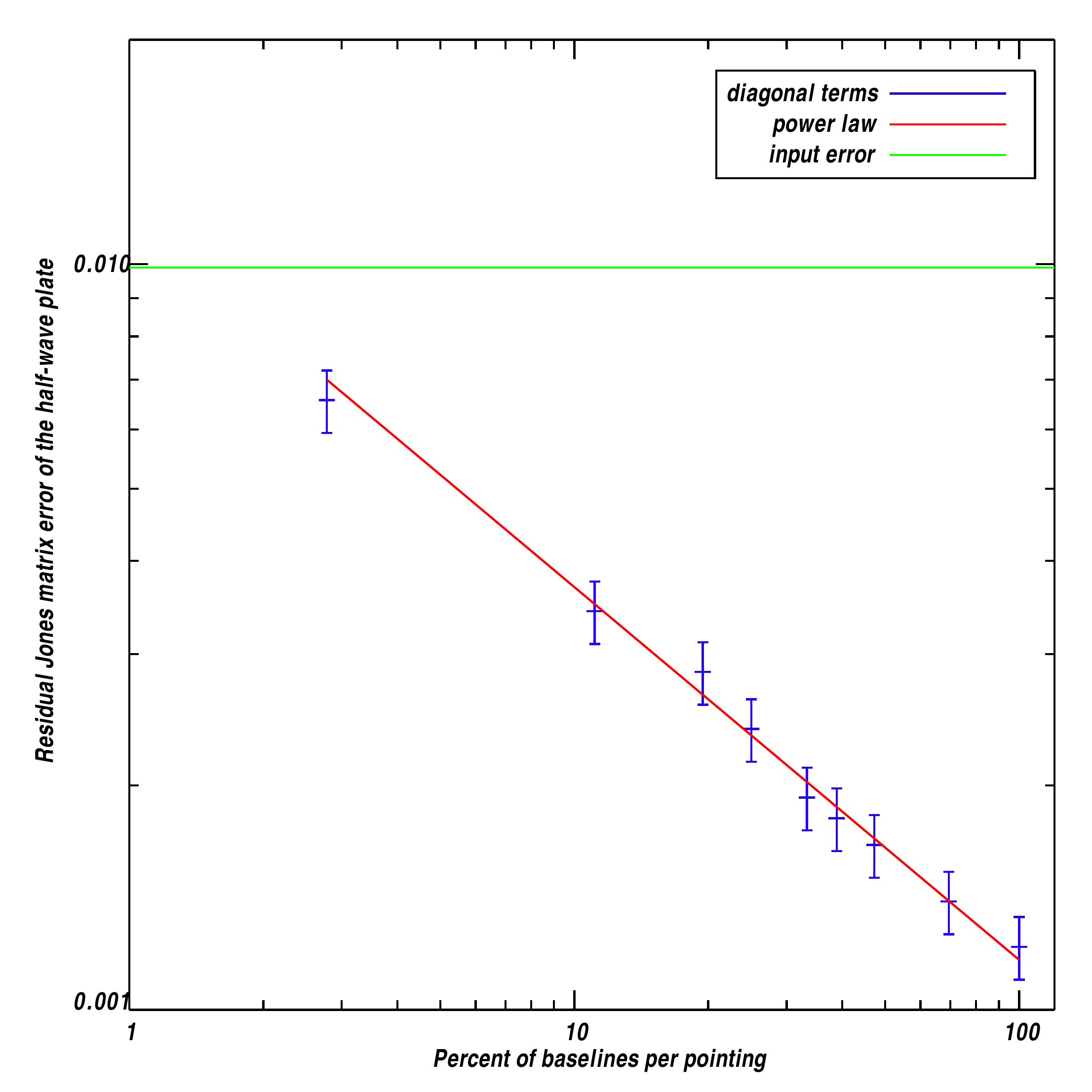}
\caption{\textit{Results of the self-calibration simulation for the diagonal terms of the Jones matrix of the half-wave plate and a time spent per baseline $t_b=1s$. This plot represents the residual error on these parameters as a function of the percent of baselines per pointing. The red line represents a power law of the shape $\sim \frac{c}{n_{bs}^\Phi}$ with c a constant, $n_{bs}$  the number of baselines per pointing and $\Phi$ the index of the power law given in Table 2. This law gives the limit of the relative accuracy that can be achieved on the systematic parameters. The green line represents the input error on the systematic effects of the half-wave plate in the simulation given in Table 1.}
\label{fig:radish}}
\end{figure}

\begin{figure}[!h] 
\centering\includegraphics[width=8cm]{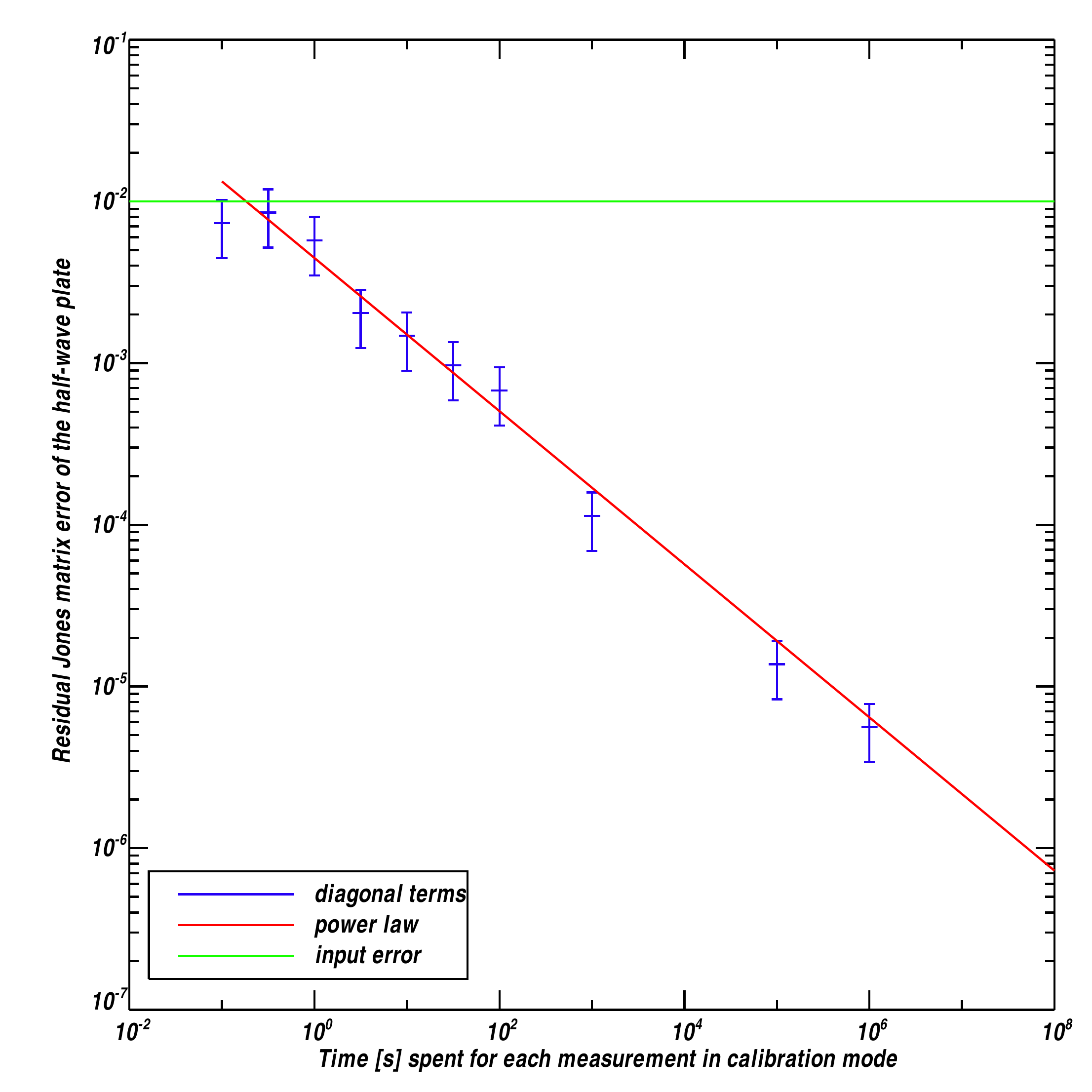}
\caption{\textit{Results of the self-calibration simulation for the diagonal terms of the Jones matrix of the half-wave plate. We show the residual error (in blue) on these parameters as a function of the measuring time spent on each baseline on the calibration mode. The red line represents a power law of the shape $\sim \frac{NET}{\sqrt{t_b}\sqrt{2}}$ with $NET$ the noise equivalent temperature of the bolometers, T the temperature of the polarized source and $t_b$ the measuring time spent on each baseline. This law gives the limit of the relative accuracy that can be achieved on the systematic parameters according to the measuring time spent per baseline. The green line represents the input error on the systematic effects of the half-wave plate in the simulation given in Table 1.}
\label{fig:radish}}
\end{figure}

The first QUBIC module will contain 400 primary horns, or 79800 baselines;  therefore we need to spend about $22$h on calibration in order to measure all the baselines during one second. 
This lapse of time could however be much reduced with a small information loss if the self-calibration procedure was not performed on all baselines. However, the accuracy on the output parameters also depends on the number of baselines per measurement as illustrated Fig.2.  It will be important to determine which is the most interesting strategy for the QUBIC instrument.

Using the law given by Eq.(\ref{eq:eq40}), one can extrapolate the result given in Table 2 to the residual error for the QUBIC instrument with 400 horns, 2$\times$1024 bolometers and 1000 pointings for two different measuring times per baseline $t_b=1$  and $t_b=100s$. The result is given in Table 3 and shows a very significant improvement on the level of the residual systematics after self-calibration, even for 1s.

\begin{table*}[ht] 
\centering
\begin{tabular}{| c | c | c | c | c | c | }
\hline
 parameters&$t_b=0s$& \multicolumn{2}{c|}{$t_b=1s$} & \multicolumn{2}{c|}{$t_b=100s$}\\
\hline
$$&$\sigma_{id-corr}$&$\sigma_{corr-rec}$&ratio&$\sigma_{corr-rec}$&ratio \\
\hline
 $\alpha_{iq}^\eta$ &  0.004&$8.48\times 10^{-5}$ &47&$1.87\times 10^{-6}$&2140\\
 $\hat{n}_p$ & 0.15&$1.41\times 10^{-3}$ &106& $3.26\times 10^{-5}$&4596\\
 $\overrightarrow{x_i}$ &  $100.\times 10^{-6}$&$5.86\times 10^{-5}$ &17&$2.27\times 10^{-8}$&4402\\
 $g_\eta(\overrightarrow{x_i})$& 0.0001&$1.36\times 10^{-6}$ &73&$1.22\times 10^{-8}$&8182\\
 $e_\eta(\overrightarrow{x_i})$ &0.0001&$1.09\times 10^{-6}$ &92&$1.20\times 10^{-8}$&8280\\
$h_\eta $ & 0.01&$1.18\times 10^{-4}$&84&$7.27\times 10^{-6}$&1375\\
 $\xi_\eta$ &0.01&$1.24\times 10^{-4}$&80&$5.81\times 10^{-6}$&1722\\ 
\hline
\end{tabular}
\caption{\textit{Results of the self-calibration simulation for each recovered parameter (given in the first column) for the QUBIC instrument with 400 horns, 2 $\times$ 1024 bolometers array, 1000 pointings and all baseline measurements. On the second column, one can see the value of the standard deviation between the ideal and corrupted parameters (without self-calibration). The third and fifth columns give the value of the standard deviation between the corrupted and reconstructed parameters (with self-calibration) for two different measuring times per baseline $t_b=1s$ and $t_b=100s$ and at total measurement time fixed to be the same for both cases.
These values are obtained by replacing in Eq.(38) the values of exponent given in Table 2 applied to the design of the QUBIC instrument. The benefit obtained after applying the self-calibration method is given in the fourth and sixth columns: this is the ratio between the value of the standard deviation of the third column and the second column for $t_b=1s$ and of the fifth and the second column for $t_b=100s$.}}
\end{table*}

\subsection{Finding limits}
Accordingly Eq.(35), in an ideal case, in the absence of systematic effects, the powers measured on the X and Y polarized focal planes after demodulation of the half-wave plate are 
\begin{equation}
\begin{pmatrix}
T_O \\
T_C \\
T_S \end{pmatrix}=
\int \int \int \begin{pmatrix}
B_{q,s}(\hat{n}_p)& 0& 0\\
0 & B_{q,s}(\hat{n}_p) & 0\\
0&0&B_{q,s}(\hat{n}_p) 
\end{pmatrix}
.\begin{pmatrix}
I(\hat{n}_p) \\
Q(\hat{n}_p) \\
U(\hat{n}_p) \end{pmatrix}
d\nu d\hat{d}_qd\hat{n}_p
  \label{eq:eq41}
\end{equation}

with $T_O$, $T_C$ and $T_S$ refer to a constant, cosine and sine terms obtained after the demodulation of the half-wave plate and the synthesized beam $B_{q,s}(\hat{n}_p)$.
The integrations are performed over the bandwidth of the instrument, over the surface of the bolometer and over the sky direction.

For a real instrument, in the case where the half-wave plate is located before the horns, leakages form Q to U and from U to Q appear. 
Using the self-calibration simulation, one can estimate the leakage from Q into U and from U into Q by calculating the standard deviation of the difference between the ideal and corrupted parameters $\sigma^{I}_{id-corr}$, $\sigma^{Q}_{id-corr}$, $\sigma^{U}_{id-corr}$, $\sigma^{QU}_{id-corr}$ and $\sigma^{UQ}_{id-corr}$ (without self-calibration) and of the difference between the corrupted and recovered parameters $\sigma^{Q}_{corr-rec}$, $\sigma^{U}_{corr-rec}$, $\sigma^{QU}_{corr-rec}$ and $\sigma^{UQ}_{corr-rec}$ (with self-calibration).
   
\vspace{0.5cm}
In this case without self-calibration, Eq.(39) becomes
\begin{equation}
\begin{pmatrix}
T_O \\
T_C \\
T_S \end{pmatrix}=
\begin{pmatrix}
\sigma^{I}_{id-corr}& 0& 0\\
0 &\sigma^{Q}_{id-corr} & \sigma^{UQ}_{id-corr}\\
0&\sigma^{U}_{id-corr}&\sigma^{QU}_{id-corr}
\end{pmatrix}
\begin{pmatrix}
I(\hat{n}_p) \\
Q(\hat{n}_p) \\
U(\hat{n}_p) \end{pmatrix}
  \label{eq:eq42}.
\end{equation} 
In this case with self-calibration, Eq.(39) becomes
\begin{equation}
\begin{pmatrix}
T_O \\
T_C \\
T_S \end{pmatrix}=
\begin{pmatrix}
\sigma^{I}_{corr-rec}& 0& 0\\
0 &\sigma^{Q}_{corr-rec} & \sigma^{UQ}_{corr-rec}\\
0&\sigma^{U}_{corr-rec}&\sigma^{QU}_{corr-rec}
\end{pmatrix}
\begin{pmatrix}
I(\hat{n}_p) \\
Q(\hat{n}_p) \\
U(\hat{n}_p) \end{pmatrix}.
  \label{eq:eq43}
\end{equation} 

With Eq.(\ref{eq:eq42}) and Eq.(\ref{eq:eq43}), one can observe that instrument errors and systematic effects induce leakage from Q to U and from U to Q but also alter the polarization amplitude.
 \vspace{0.5cm} 

It is interesting to focus on the B-mode power spectrum in order to estimate the E-B mixing and to constrain the E-mode leakage in the B-mode power spectrum. 

In general, one can define the vector $\Delta \overrightarrow{S}$ which defines the errors on Stokes parameters
 \begin{equation}
\Delta \overrightarrow{S} = \begin{pmatrix}\Delta_{II}&\Delta_{IQ}&\Delta_{IU}\\
\Delta_{QI}&\Delta_{QQ}&\Delta_{QU}\\
\Delta_{UI}&\Delta_{UQ}&\Delta_{UU}\\
\end{pmatrix}\overrightarrow{S}
  \label{eq:eq44}
 \end{equation}
where $\Delta \overrightarrow{S}= \overrightarrow{S}^{rec}-\overrightarrow{S}^{corr}$.
 
where $\overrightarrow{S}^{corr}$ is the vector of measured Stokes parameters and $\overrightarrow{S}^{rec}$ is the vector of the Stokes parameters obtained with the self-calibration method. 
For the QUBIC instrument, there is no leakage of I to Q and U because the half-wave plate is in front of the instrument and so, Eq.(44) becomes
 \begin{equation}
\Delta \overrightarrow{S} = \begin{pmatrix}0&0&0\\
0&\Delta_{QQ}&\Delta_{QU}\\
0&\Delta_{UQ}&\Delta_{UU}\\
\end{pmatrix}\overrightarrow{S}
  \label{eq:eq45}
 \end{equation}
where $\Delta_{QQ}$, $\Delta_{QU}$, $\Delta_{UQ}$ and $\Delta_{UU}$, the errors between the synthesized beam without the self-calibration method and the one after applying the self-calibration method.
%Table 5 gives the values of $\Delta_{QQ}$, $\Delta_{QU}$, $\Delta_{UQ}$ and $\Delta_{UU}$ for X and Y focal planes. It shows that the value of the rms decreases when the self-calibration is applying.

Following Eq.(\ref{eq:eq45}), we can define the error matrix on the Q and U Stokes parameters 
 \begin{equation}
\mathbf{M_s} = \begin{pmatrix}
1+\epsilon&\rho\\
-\rho&1+\epsilon\\
\end{pmatrix}
  \label{eq:eq46}
 \end{equation} 
where the complex term $\epsilon$ changes the amplitude of polarization and $\rho$ mixes both the Q and U Stokes parameters. The justification of this matrix is given in Appendix B.

To go further, one can estimate the leakage from E to B-mode and give a constraint on the value of the tensor-to-scalar ratio r.

An error in diagonal terms $\Delta_{QQ}$ and $\Delta_{UU}$ will affect the amplitude of the E and B-mode power spectrum. The non diagonal terms $\Delta_{QU}$ and $\Delta_{UQ}$ results in a leakage from the E to B-mode power spectrum (or the B to E-mode power spectrum). 

To have a constraint on the B-mode, as the E-mode is far above that of the B-mode in amplitude, one can use the equation derived by (C. Rosset et al. 2010)
  \begin{equation}
  \Delta  Cl^{BB}=Cl^{BB, meas}-Cl^{BB}=2\epsilon Cl^{BB}+\rho^2  Cl^{EE}
    \label{eq:eq47}
   \end{equation}
where $\Delta  Cl^{BB}$ is the global error on the B-mode power spectrum $Cl^{BB}$ and $Cl^{EE}$ is the E-mode power spectrum.
One can refer to Appendix B for explicit details of this equation.
It shows that the uncertainty on parameter $\rho$ must be lower than 0.5$\%$ to have a leakage from E to B-mode lower than 10$\%$ for a tensor-to-scalar ratio of r=0.01 for $l<$100.

From Eq.(\ref{eq:eq47}), one can estimate the leakage from the E-mode into the B-mode power spectrum given by the term $\rho^2  Cl^{EE}$. Fig.4 represents the error on the B-mode power spectrum as a function of the multipoles. The colored solid lines represent the leakage after applying the self-calibration method for different measuring times per baseline from $t_b=$1s to 10000s. The dashed line represents the error on the B-mode power spectrum without self-calibration method. The black lines are the initial B-mode spectrums for $r = 10^{-1}$, $r=10^{-2}$ and $r=10^{-3}$.

The leakage from the E-mode into the B-mode is therefore significantly reduced by applying the self-calibration procedure, even with a modest 1s per baseline (corresponding to a full day dedicated to self-calibration). The leakage can be further reduced by spending more time on self-calibration.

\begin{figure}[!h]
\centering\includegraphics[width=7.5cm]{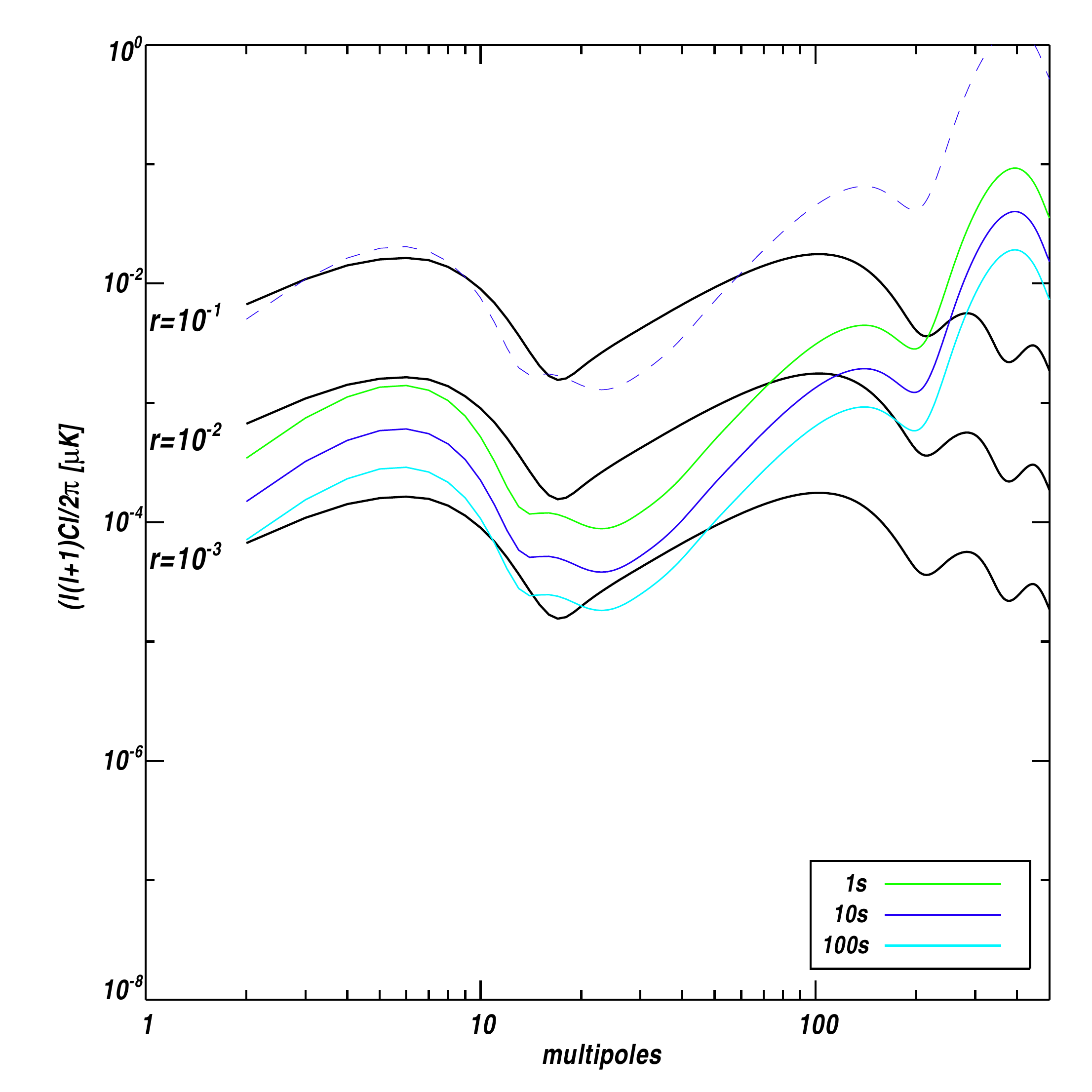}
\caption{\textit{$\Delta C_l$ due to leakage from E-mode for different times on measurements per baseline $t_b=$1s, 10s, 100s obtained with the self-calibration method (solid lines) and without the self-calibration method (dashed blue line) for the QUBIC instrument. The black lines are the primordial B-mode spectrums for $r=10^{-1}$, $r=10^{-2}$ and $r=10^{-3}$.}
\label{fig:radish}}
\end{figure}

\section{Conclusion}
Bolometric interferometry differs considerably from standard radio interferometry in the sense that its primary goal is not to reach a good angular resolution but to achieve high statistical sensitivity and good control of systematic effects. In this perspective, redundancy turns out to be the crucial property to fulfill these two objectives, as shown in (Charlassier et al. 2009) and in the present article.

In this paper, we have shown that with a polarized calibration source and the use of the successive observation of this source with all pairs of horns of the interferometer (self-calibration), one can have low and controllable instrumental systematic effects. Redundant baselines should give the same signal if they are free from systematics. By modeling the instrument systematics with a set of parameters (Jones matrices, location of the horns, beams), one can use the measurements of the different baselines to solve a non-linear system that allows to determine the systematic effects parameters with an accuracy that, besides the correctness of the modelling, is only limited by the photon noise, and hence by the time spent on self-calibration. The more horns and bolometers in the array, the more efficient the self-calibration procedure.

The resolution of the system is CPU-intensive for large bolometric interferometers and should be implemented on massively parallel computers in the future. Using simulations with various horns and bolometer arrays of moderate sizes, we have obtained a scaling law that allows us to extrapolate the accuracy of self-calibration to the QUBIC instrument with 400 horns, 2x1024 bolometer arrays and 1000 pointing directions towards the calibration source. We find that with a few seconds per baseline (corresponding to a few days spent on self-calibration), the knowledge of the instrumental systematic effects parameters can be improved by at least two orders of magnitude, allowing to minimize the leakage from E into B polarization down to a tolerable level. This can be improved by spending more time on self-calibration.

The idea of developing bolometric interferometry was motivated by bringing together the imager exquisite sensitivity allowed by bolometer arrays and the ability to handle instrumental systematic effects allowed by interferometers. Bolometric interferometers have been shown to have a similar sensitivity to imagers (Hamilton 2008 ; The QUBIC collaboration 2010) while we show in the present article that the self calibration allows to achieve an excellent handling of systematic effects which has no equivalent with an imager.

 \begin{acknowledgements}
The authors are grateful to the QUBIC collaboration and Craig Markwardt for MPFIT. They would like to thank Gael Roudier for his help. This work was supported by Agence Nationale de la Recherche (ANR), Centre National de la Recherche Scientifique (CNRS) and la r\'egion d'Ile de France.
\end{acknowledgements}

 \appendix

\section{Measuring the bolometer power for two opened horns is equivalent to measuring the bolometer power when all horns are open except the horn i and j}

The power collected by the bolometer $q$ for the opened horns i and j without polarization is
  \begin{equation}
  S_{ij}=C_i+C_j+2Re(\alpha_{iq}\alpha_{jq}^*\beta_{iq}\beta_{iq}^*)
   \end{equation}
which can be written as:
   \begin{equation}
 S_{ij}=|p_i|^2+|p_j|^2+2Re(p_ip_j^*).
   \end{equation}

The total power measured for all the baselines can be expressed as
  \begin{equation}
S_{tot}=|\sum_m p_m|^2=\sum_m |p_m|^2+\sum_{m\neq l}2Re(p_mp_l^*) .
   \end{equation}
The power measured by a bolometer $q$ for all horns opened except the horn i is
     \begin{equation}
C_{-i}=|\sum_{j\neq i} p_j|^2=S_{tot}-C_{i}-\sum_{k\neq i} 2Re(p_ip_k^*).
   \end{equation}
The power measured by a bolometer $q$ for all horns opened except the horn j is
     \begin{equation}
C_{-j}=|\sum_{i\neq j} p_j|^2=S_{tot}-C_{j}-\sum_{k\neq j} 2Re(p_jp_k^*).
  \end{equation}
The power measured by a bolometer q for all baselines opened except the baseline formed by the horns i and j is
  \begin{equation}
S_{-ij}=S_{tot}-C_i-C_j-\sum_{k\neq i}2Re(p_ip_k^*)-\sum_{k\neq j}2Re(p_jp_k^*)+2Re(p_ip_j^*)
  \end{equation}

with
\begin{multline*}
2Re(p_ip_j^*)=S_{tot}+S_{-ij}-S_{tot}+C_i+\sum_{k\neq i}2Re(p_ip_k^*)-S_{tot}+ \nonumber \\
C_j+\sum_{k\neq j}2Re(p_jp_k^*).
\end{multline*}

 Finally, one can find
 \begin{equation}
2Re(p_ip_j^*)=S_{tot}+S_{-ij}-S_{tot}-C_{-i}-C_{-j}.
  \end{equation}
 So :
  \begin{equation}
S_{-ij}=C_{-i}+C_{-j}+2Re(p_ip_j^*).
  \end{equation}

 \section{Error on E and B-modes power spectrum}
 
 One can define the Stokes parameters in spin-2 spherical harmonics base
 \begin{equation}
 Q(\hat{n})\pm iU(\hat{n})=\sum_{lm}a_{\pm 2lm \pm 2}Y_{lm}(\hat{n})
\end{equation}
where Q and U are defined at each direction $\hat{n}$.

It is convenient to introduce the linear combinations
\begin{equation}
a^E_{lm}=-\frac{a_{2lm}+a_{-2lm}}{2}  
\end{equation}
\begin{equation}
a^B_{lm}=i\frac{a_{2lm}-a_{-2lm}}{2}.
\end{equation}
One can define the two scalar fields
 \begin{equation}
 E(\hat{n})=\sum^{\infty}_{l=0} \sum^{l}_{m=-l}  a^E_{lm}Y_{lm}(\hat{n})
\end{equation}
 \begin{equation}
 B(\hat{n})=\sum^{\infty}_{l=0} \sum^{l}_{m=-l}  a^B_{lm}Y_{lm}(\hat{n}).
\end{equation}

Using the coefficients $ a^E_{lm}$ and $ a^B_{lm}$, one can construct the angular power spectrum $C_l^{EE}$ and $C_l^{BB}$ as
 \begin{equation}
C_l^{EE}=< \mid a^E_{lm}\mid^2  >
\end{equation}
 \begin{equation}
 C_l^{BB}=< \mid a^B_{lm}\mid^2  >.
\end{equation}

One can express the Q and U Stokes parameters as a function of the coefficients $ a^E_{lm}$ and $ a^B_{lm}$
 \begin{equation}
Q(\hat{n})=\frac{1}{2} \sum_{l,m}[(ia^B_{lm}-a^E_{lm})_{-2}Y_{lm}(\hat{n})+(-ia^B_{lm}-a^E_{lm})_{+2}Y_{lm}(\hat{n})]
 \end{equation}
  \begin{equation}
 U(\hat{n})=\frac{1}{2}\sum_{l,m}[i(ia^B_{lm}-a^E_{lm})_{-2}Y_{lm}(\hat{n})+(-ia^B_{lm}-a^E_{lm})_{+2}Y_{lm}(\hat{n})].
 \end{equation}

Using Eq.(B.4), (B.5), (B.8) and (B.9), one can express the coefficients $ a^E_{lm}$ and $ a^B_{lm}$ as a function of the Stokes parameters Q and U
 \begin{equation}
 \begin{pmatrix}
a^E_{lm} \\
a^B_{lm}
\end{pmatrix}
=  \int \mathbf{M_{alm}}
\begin{pmatrix}
Q(\hat{n}) \\
U(\hat{n}) 
\end{pmatrix}
d\Omega
  \end{equation}
where the integration is taken over the whole sky and

 \begin{equation}
 \mathbf{M_{alm}}=
  \begin{pmatrix}
-(_{+2}Y_{lm}(\hat{n})+_{-2}Y_{lm}(\hat{n})) & -i(_{+2}Y_{lm}(\hat{n})-_{-2}Y_{lm}(\hat{n})) \\
i(_{+2}Y_{lm}(\hat{n})-_{-2}Y_{lm}(\hat{n}) & -(_{+2}Y_{lm}(\hat{n})+_{-2}Y_{lm}(\hat{n}))\end{pmatrix}.
  \end{equation}

In the case of a bolometric interferometer, globar errors on synthesized beam will affect the amplitude of polarization and mix the Q and U Stokes parameters.

To model systematic errors, one can introduce a Jones matrix which describes the propagation of radiation through a receiver
\begin{equation}
\overrightarrow{E_r}=\mathbf{J}\overrightarrow{E}=
\begin{pmatrix}
1-g_{x} & e_{x}  \\
e_{y} & 1-g_{y} 
\end{pmatrix}
\overrightarrow{E}
\end{equation}

where the gain $g_\eta$ and the leakage $e_\eta$ are complex values.

From this relation, one can construct the Mueller matrix $\mathbf{M}$ which tells us how the Stokes vector $\overrightarrow{S}$ transforms

\begin{equation}
\overrightarrow{S}_r=\mathbf{A}(\mathbf{J} \otimes  \mathbf{J}^*)\mathbf{A}^{-1}\overrightarrow{S}=\mathbf{M}\overrightarrow{S}
\label{eq:eq13}
\end{equation}

where  $\mathbf{A}=
\begin{bmatrix}
1 & 0 & 0 & 1\\
1 & 0 & 0 & -1 \\
0 & 1 & 1 & 0\\
0 & i & -i & 0\\
\end{bmatrix}
$
and $\overrightarrow{S}_r$ is the outgoing Stokes vector and $\mathbf{M}=\mathbf{A}(\mathbf{J} \otimes  \mathbf{J}^*)\mathbf{A}^{-1}$ is the resulting error matrix.

\vspace{0.5cm}

We are only interested on the Q and U Stokes parameters. In this case, the error matrix $ \mathbf{M}$ becomes
 \begin{equation}
\mathbf{M} =
\begin{pmatrix}
M_{11} & M_{12}   \\
M_{21} & M_{22} 
\end{pmatrix}.
\end{equation}

The first order of this matrix $\mathbf{M}$ is

$M_{11}=1+g_x+g_y+g_x^*+g_y^*$

$M_{21}=e_y+e_y^*-e_x-e_x^*$

$M_{12}=e_x+e_x^*-e_y-e_y^*$

$M_{22}=1+g_x+g_y+g_x^*+g_y^*$.

\vspace{0.5cm}

One can generalize this error matrix for Jones matrix of any component and rewrite it as

 \begin{equation}
\mathbf{M_s} = \begin{pmatrix}
1+\epsilon&\rho\\
-\rho&1+\epsilon\\
\end{pmatrix}
 \end{equation}
where the complex term $\epsilon$ describes the error of the amplitude of polarization and the complex term $\rho$ mixes the Q and U Stokes parameters.

In this case, Eq.(B.10) becomes
 \begin{equation}
 \begin{pmatrix}
a^{E,meas}_{lm} \\
a^{B,meas}_{lm}
\end{pmatrix}
=  \mathbf{M_{alm}}\mathbf{M_s}
\begin{pmatrix}
Q(\hat{n}) \\
U(\hat{n})
\end{pmatrix}
  \end{equation}
where the coefficients $a^{E,meas}_{lm} $ and $a^{B,meas}_{lm} $ include systematic effects.
 
In terms of power spectra, an error in polarization amplitude will affect the amplitude of the the E and B-modes power spectra and  an error which mixes the Q and U Stokes parameters leads to a leakage from the E to B-mode (and from the B to E-mode).

One can easily find that with systematic effects, the coefficients $ a^E_{lm}$ and $ a^B_{lm}$ can be expressed as
  \begin{equation}
  a^{E,meas}_{lm}=  a^E_{lm}+\epsilon a^E_{lm}-\rho a^B_{lm}
   \end{equation}
   \begin{equation}
  a^{B,meas}_{lm}=  a^B_{lm}+\epsilon a^B_{lm}+\rho a^E_{lm}
   \end{equation}
where $\epsilon$ is the error in amplitude and $\rho$ the error of polarization leakage.
   Using Eq.(B.6) and Eq.(B.7), one can obtain
   \begin{equation}
  Cl^{EE,meas}=  Cl^{EE}+2\epsilon Cl^{EE}+\rho^2 Cl^{BB}
   \end{equation}
   \begin{equation}
 Cl^{BB, meas}=   Cl^{BB}+2\epsilon Cl^{BB}+\rho^2  Cl^{EE}
   \end{equation}
where $Cl^{EE,meas}$ and $Cl^{BB, meas}$ are the E and B-mode power spectra including systematic effects, $Cl^{EE}$ and $Cl^{BB}$ the input power spectra,  the complex term $\epsilon$ describes the error of amplitude of the B-mode power spectrum and the complex term $\rho$ results in a leakage from the E to B-mode power spectrum.

To focus on the B-mode, one can define the error on $Cl^{BB}$ power spectrum
   \begin{equation}
  \Delta  Cl^{BB}=Cl^{BB, meas}-Cl^{BB}=2\epsilon Cl^{BB}+\rho^2  Cl^{EE}.
   \end{equation}


\begin{thebibliography}{}

\bibitem{la}Bock, J., Church, S., Devlin, M., et al. 2006, ArXiv Astrophysics e-prints
\bibitem{ja} Bunn, E. F. 2007, Phys. Rev. D, 75, 083517
\bibitem{aa}Charlassier R. 2010, PhD, University Paris-Diderot
\bibitem{ja}Charlassier, R., Bunn, E. F., Hamilton, J., Kaplan, J., \& Malu, S. 2010, \aap, 514, A37+
\bibitem{fa}Charlassier, R., Hamilton, J., Br\'eelle, E., et al. 2009, \aap, 497, 963
\bibitem{ha}Hamilton , J.-C. and Charlassier , R. and Cressiot , C. and Kaplan , J. and Piat , M. and Rosset , C. 2008, \aap, 491, 923
\bibitem{ta}Hu, W., Hedman, M. M., \& Zaldarriaga, M. 2003, Phys. Rev. D, 67, 043004
\bibitem{ga}Hyland, P., Follin, B., \& Bunn, E. F. 2009, MNRAS, 393, 53
\bibitem{dasi}Kovac, J., Leitch, E. M., Pryke, C., et al. 2002, Nature, 420, 772
\bibitem{ba}Liu, A.,Tegmark, M., Morrison, S., Lutomirski, A., \& Zaldarriaga, M. 2010, ArXiv e-prints
\bibitem{ba}Markwardt, C. B. 2009, in Astronomical Society of the Pacific Conference Series, Vol. 411, Astronomical Society of the Pacific Conference Series, ed. D. A. Bohlender, D. Durand, \& P. Dowler, 251Ð+
\bibitem{ma}Noordam, J. E. \& de Bruyn, A. G. 1982, Nature, 299, 597
\bibitem{ka}O'Dea, D., Challinor, A., \& Johnson, B.R. 2007, MNRAS, 376, 1767
\bibitem{ma}Pearson, T. J. \& Readhead, A. C. S. 1984, ARA\&A, 22, 97
\bibitem{cbi}Readhead, A.C.S., Myers, S.T., Pearson, T.J., {\it et al.} 2004, {\em Science}, 306, 836
\bibitem{ca}Rosset, C., Tristram, N., Ponthieu, N., et al. 2010, \aap, 520, A13 
\bibitem{da}Tegmark, M. \& Zaldarriaga, M. 2009a, Phys. Rev. D, 79, 083530
\bibitem{ea}Tegmark, M. \& Zaldarriaga, M. 2009b,  ArXiv e-prints
\bibitem{ja}The QUBIC Collaboration 2010,  Astroparticle Physics 34
\bibitem{ya}Timbie, P. T., Tucker, G. S., Ade, P. A. R., et al. 2006, New Astronomy Review, 50, 999
\bibitem{ma}Tucker, G. S., Kim, J., Timbie, P., et al 2003, New Astronomy Review, 47, 1173 
\bibitem{cbi}Wieringa, M. 1991, in Astronomical Society of the Pacific Conference Series, Vol. 19, IAU Colloq. 131: Radio Interferometry. Theory, Techniques, and Applications, ed. T. J. Cornwell \& R. A. Perley, 192Ð196
\end{thebibliography}
\end{document}